\documentclass[12pt,doublespace]{article}
\begin{document}
\begin{center}
MULTIPLEX CHAOS SYNCHRONIZATION IN SEMICONDUCTOR LASERS WITH MULTIPLE OPTOELECTRONIC FEEDBACKS \\
E.M.Shahverdiev $^{1,2,*}$ and K.A.Shore $^{1}$\\
$^{1}$School of Electronic Engineering,Bangor University, Dean St.,Bangor, LL57 1UT, Wales, UK\\
$^{2}$Institute of Physics, H.Javid Avenue,33, Baku, AZ1143, Azerbaijan\\
$^{*}$e-mail:shahverdiev@physics.ab.az\\
 tel: 994-12-537-21-49\\
fax:994-12-539-59-61\\
~\\
ABSTRACT
\end{center}
Secure chaos based multiplex communication system scheme is proposed utilizing globally coupled semiconductor lasers with multiple variable time delay optoelectronic feedbacks.
~\\
PACS number(s):42.55.Px, 42.65.Sf, 05.45.Xt, 42.60.Mi,05.45.Gg, 05.45.Vx\\
\begin{center}
I. INTRODUCTION
\end{center}
\indent In recent years due to its fundamental and applied interests chaos synchronization has been the subject of extensive studies [1-2]. Chaos-based communications is emerging as an alternative technique to improve security in such systems, especially after the recent field demonstration using a metropolitan fibre network [3]. For message decoding in such schemes one has to be able to synchronize the transmitter and receiver lasers [1-4].\\
In [5] the authors have presented a multiplexed chaos communication scheme utilizing globally coupled multiple transmitter (drive) and  receiver(response) chaotic systems. The transmitter and the receiver were comprised  of multiple non-identical chaotic systems. Each of the transmitter subsystems was synchronized to the identical receiver subsystem  by the coupling signal which was at the same time an information signal. The authors of [5] have suggested that an increase of the security of such a communication scheme  can be achieved by making a scalar global coupling signal truely random by construction. On the one hand, this scheme can guarantee the possiblity of enhanced security because of the stochasticity of the carrier (coupling signal). On the other hand, the chaotic nature of the underlying generators could also provide enhanced security.\\ 
However, in [5] underlying chaos generator was the Lorenz system- the system of three ordinary differential equations (ODE) which is a very popular model in nonlinear chaos dynamics [6], albeit with limited practical interest. Moreover as proven in [7], ODE chaos based communuication systems could be very vulnerable from the security point of view. \\
\indent Initially it was suggested that, from the security point of view time-delayed systems [8] exhibit intriguing characteristics with increasing time delay: despite a small number of dynamical variables the number of positive Lyapunov exponents increase with the delay time, and the system exhibits hyperchaos dynamics [9]. Moreover, as emphasized and demonstrated in [10] multiple time delay systems are capable of offering even more complexity than a single time-delayed systems. However, it was found later in [11] that communications based on chaotic time delayed systems is also vulnerable as the delay times can be exposed by several methods e.g.  the autocorrelation function [12] technique.\\
\indent As established in [11,13], modulation of time delays can erase the signatures of time delays in the autocorrelation function, thus providing more secure chaos-based communication systems.\\
\indent In this work we propose secure chaos-based multiplexed communication scheme based on variable multiple time-delayed semiconductor lasers with  optoelectronic feedbacks. The transmitter (drive) system comprises non-identical pairs (subsystems) of time delayed lasers. The receiver (response) system is identical to the transmitter. The coupling between the transmitter and the receiver lasers is global, i.e. the output of the transmitter system is added to both the receiver system and the transmitter system itself. We also present the case when the coupling signal between the transmitter and the reciever systems can play the role of the information signal in transmitting a binary message, as suggested in [4].\\
\begin{center}
II. SYSTEM MODEL
\end{center}
\indent The main constituent of the proposed multiplex communication system is a semiconductor laser with several variable time delay optoelectronic feedbacks. As a proof of principle and to be more specific, consideration is given to transmitter and receiver semiconductor lasers subject to optoelectronic feedbacks with two variable time delays and both comprise two subsystems. For the case of optoelectronic feedback 
the optical power emitted by the laser is divided into two parts, detected, amplified, and added to their own injection current (figure 1). Multiplex chaos communication scheme proposed here will comprise :(a) the transmitter system consisting of two  non-identical semiconductor lasers with optoelectronic feedbacks;(b)the receiver system also consisting of two no-identical lasers with optoelectronic feedbacks. Consideration will be given to synchronization  between the transmitter subsystem lasers identical to the respective receiver subsystem lasers. The coupling signal is chosen to be the output of the transmitter subsystem lasers. In transmitting a binary message, the output of the one of  subsystem  transmitter lasers will be chosen if for example binary bit ''0'' is to be transmitted. The other subsystem transmitter laser output will be chosen if binary bit ''1'' is to be transmitted. It is noted that in each case of binary transmission the coupling signal is also fed into the transmitter subsystem lasers.\\
The dynamics of the double time delay constituent transmitter subsystem lasers are governed by the following systems
\begin{equation}
\frac{dS_{1}}{dt}=(\Gamma g_{1} -\gamma_{c})S_{1}
\end{equation}
\begin{equation}
\frac{dN_{1}}{dt}=I_{1} - \gamma_{s1} N_{1} - g_{1}S_{1} + \gamma_{c}(k_{1} S_{1}(t-\tau_{1}) + k_{2} S_{1}(t-\tau_{2}) + KS_{CS}(t-\tau_{3}))
\end{equation}
\begin{equation}
\frac{dS_{2}}{dt}=(\Gamma g_{2} -\gamma_{c})S_{2}
\end{equation}
\begin{equation}
\frac{dN_{2}}{dt}=I_{2} - \gamma_{s2} N_{2} - g_{2}S_{2} + \gamma_{c}(k_{3} S_{2}(t-\tau_{1}) + k_{4} S_{2}(t-\tau_{2}) + KS_{CS}(t-\tau_{3})) 
\end{equation}
where subindicies 1,2 distinguish between the transmitter subsystems;$S_{1,2}$ is the photon density; $N_{1,2}$ is the carrier density;
$g_{1,2}$ is the material gain;$\gamma_{c}$ is the cavity decay rate;$\gamma_{s}$ is the carrier relaxation rate;$\Gamma$ is the confinement factor of the laser waveguide.$I_{1,2}$ is the injection current (in units of the electron charge);
$k_{1,2}$ and $k_{3,4}$ are the feedback rates for the transmitter subsystems,respectively; $\tau_{1,2}$
are the feedback delay times;$\tau_{3}$ is the delay time for the coupling signal to be fed into the transmitter subsystem lasers. The term $KS_{CS}(t-\tau_{3})$ in the right-hand side of Eqs.(2) and (4) is the coupling signal (CS) with coupling rate $K$.The dynamics of the double time delay constituent receiver subsystem lasers are identical to those for the respective transmitter subsystems. The same coupling term $KS_{CS}(t-\tau_{3})$ 
is added to both subsystems of the receiver laser. In figure 2 a schematic diagram of the multiplex chaos communication scheme is presented.\\
In a wide operation range the material gain $g$ can be expanded as 
\begin{equation}
g\approx g_{0} + g_{n}(N-N_{0}) + g_{p}(S-S_{0}),
\end{equation}
where $g_{0}=\gamma_{c}/\Gamma $ is the material gain at the solitary threshold; $g_{n}=\partial g/\partial N >0$ is the differential gain parameter; 
$g_{p}=\partial g/\partial S <0$ is the nonlinear gain parameter; $N_{0}$ is the carrier density at threshold; $S_{0}$ is the free-running intracavity photon density when the lasers are decoupled;the parameters $g_{n}$ and $g_{p}$ are taken to be approximately constant.\\
In this paper we will mainly consider variable time delays $\tau_{1}(t),\tau_{2}(t)$, and $\tau_{3}(t);$
we choose the following form for the modulation of time delays:
\begin{equation}
\tau_{1,2}=\tau_{01,02} + S^{f}_{1}(t) \tau_{a1,a2}\sin(\omega_{1,2}t)
\end{equation}
are the variable feedback delay times.  The variable time of flight between the transmitter laser and receiver laser,
\begin{equation}
\tau_{3}=\tau_{03} + S^{f}_{1}(t) \tau_{a3}\sin(\omega_{3}t)
\end{equation}
which is also the delay time for the coupling signal to be fed into the transmitter subsystems; $\tau_{01,02,03}$ are mean time delays, $\tau_{a1,a2}$ are the amplitude,$\omega_{1,2,3}/2\pi$ are the frequency of the modulations; $S^{f}_{1}(t)$ is the output power of the transmitter subsystem laser, Eqs.(1-2) for constant time delays.\\
\begin{center}
III. NUMERICAL SIMULATIONS AND DISCUSSIONS
\end{center}
In the numerical simulations we use typical values for the internal parameters of the transmitter subsystem lasers:
$\gamma_{c}=2 ps,\gamma_{s1}=2 ns, \Gamma =0.3, g_{0}=7\times 10^{13},g_{n}=10^4,g_{p}=10^4, N_{0}=1.7\times 10^{8},S_{0}=5\times 10^{6}, 
I_{1}=3.4\times 10^{17}$, see, e.g.[14]. The parameters of the receiver subsystem lasers are chosen to be identical to those of the respective transmitter subsystem lasers. The subsystem lasers are made non-identical by the different feedback levels.\\
\indent Before studying the multiplex synchronization between the transmitter and receiver lasers with variable time delays we investigate the autocorrelation coefficient for the output of the constituent transmitter laser for both constant and variable time delays. The autocorrelation coefficient is a measure of how well a signal matches a time shifted version of itself and is a special case of the cross-correlation coefficient [12]
\begin{equation}
C(\Delta t)= \frac{<(x(t) - <x>)(y(t+\Delta t) - <y>)>}{\sqrt{<(x(t) - <x>)^2><(y(t+ \Delta t) - <y>)^2>}},
\end{equation}
where x and y  are the outputs of the interacting laser systems; the brackets $<.>$  represent the time average;  $\Delta t$  is a time shift between laser outputs. This coefficient indicates the quality of synchronization: C=0 implies no synchronization; C=1 means perfect synchronization.\\
\indent Figure 3 demonstrates the autocorrelation coefficient ($C_{A}\equiv C(x=y)$) for the output of transmitter subsystem laser, Eqs. (1-2) for constant time delays,respectively, i.e. for $\omega_{1}=\omega_{2}=0, k_{1}=0.80, k_{2}=0.70, \tau_{01}=4X10^{-9}s,\tau_{02}=6X10^{-9}s.$ It is clearly seen that time delays can be easily recovered from both the autocorrelation coefficient, which exhibits extrema at the time delays or their multiples and combinations.\\
\hspace*{0.5cm} Next let us consider the variable time delay scenario. In investigating the behavior of the autocorrelation coefficient we have experimented with different types of variable time delays, among them:(a)sinusoidal modulations:$\tau_{1,2}=\tau_{01,02} + \tau_{a1,a2}\sin(\omega_{1,2}t);$ (b)chaotic modulations:$\tau_{1,2}=\tau_{01,02} + \tau_{a1,a2} S^{f}_{1}(t);$and (c)combined chaotic and sinusoidal modulations $\tau_{1,2}=\tau_{01,02} + S^{f}_{1}(t)\tau_{a1,a2}\sin(\omega_{1,2}t).$ Extensive numerical simulations have established that erasure of the signatures of time delays in the autocorrelation coefficient is best achieved for combined chaotic and sinusoidal modulations of $\tau(t).$
Figure 4 shows the autocorrelation coefficient  of the transmitter subsystem laser output, Eqs.(1-2) for combined sinusoidal and chaotic time delays for 
$\tau_{1}(t)= (4X10^{-9}  +  4X10^{-17}S^{f}_{1}(t)\sin(2X10^{6}t))$s and $\tau_{2}(t)= (6X10^{-9} + 4X10^{-17}S^{f}_{1}(t)\sin(2X10^{6}t))$s
with the rest of parameters as for figure 3. Thus, modulation of the delay times gives rise to the loss of their signature in the autocorrelation coefficient, and therefore can improve the security of chaos based communication systems.\\
As mentioned above, in the multiplex chaos based communication scheme proposed in this work we will consider synchronization between the identical  transmitter and receiver subsystem lasers. First we present the results of the numerical simulations, when the output (the intensity) of the laser subsystem, Eqs.(1-2) is added to both the transmitter laser subsystems and to both the receiver laser subsystems. In practice, as emphasized above in binary message transmission e.g. the output $S_{1}$($S_{2}$)can be used for the transmission of ''0''(''1''  )bits. Figure 5(a) shows the dynamics of one of the transmitter subsystem laser intensity $S^{t}_{1}(t)$ for such a case. Figure 5(b) demonstrates the synchronization error dynamics between the receiver subsystem $S^{r}_{1}$ and the transmitter subsystem $S^{t}_{1}$ intensities $S^{r}_{1}-S^{t}_{1}$. High quality synchronization is also observed (not presented here) between the second subsystem transmitter laser and the second receiver subsystem laser. The highest quality synchronization (C=1) was also achieved  between the respective subsystems of the transmitter and receiver in the case of the coupling signal $S_{2}.$ \\
\indent We have also experimented with other types of the coupling signal. Figure 6(a)) presents the dynamics of the laser intensity $S^{t}_{2}(t)$ for the case of $S_{1}(t-\tau_{3})- S_{2}(t-\tau_{3})$ coupling. Figure 6(b) shows the dynamics of the error signal $ S^{t}_{2}(t)-S^{r}_{2}(t)$ dynamics for this case. Again the highest quality synchronization between the corresponding subsystem lasers of the transmitter and receiver laser is achieved. It is noted that the high quality of chaos synchronization observed in all cases is an important requirement in chaos-based communication systems. It should also be emphasized that numerical simulations revealed that synchronization is very robust to parameter mismatches (5-7$\%$), which is of certain importance from the practical point of view.\\ 
\begin{center}
IV.CONCLUSIONS
\end{center}
\indent In conclusion, we have demonstrated that the time-delay signature is eliminated from the laser output autocorrelation  in systems with modulated optoelectronic feedbacks. We have also described multiplex chaos synchronization in globally coupled variable multiple time delay lasers with optoelectronic feedbacks. The results of the paper provide the basis for the use of the optoelectronic feedback lasers with multiple variable time delays in enhanced security chaos-based high-speed communication systems.\\
\begin{center}
V.ACKNOWLEDGEMENTS
\end{center}
This research was supported by a Marie Curie Action within the 6th European Community Framework Programme Contract.\\
\begin{center}
Figure Captions
\end{center}
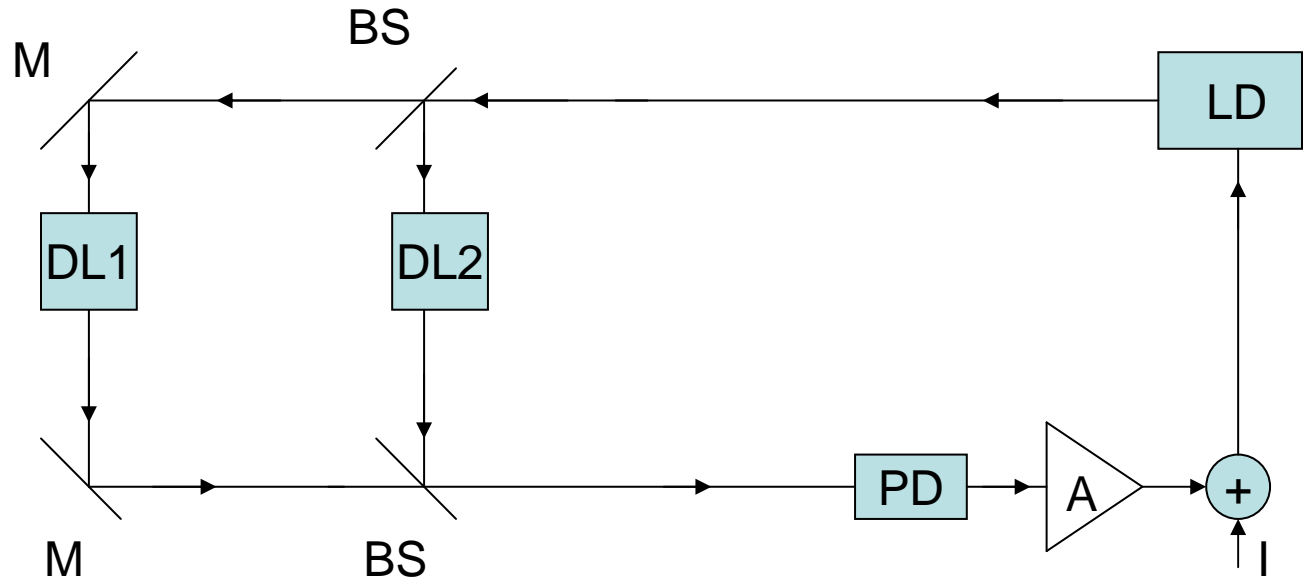
\noindent FIG.1. Schematic arrangement for the lasers with double optoelectronic feedback.LD:Laser diode. PD:Photodetector.BS:Beamsplitter.M:Mirror. DL:Delay lines. A:Amplifier.I:Injection current. The output of the laser is split by a beamsplitters and directed along different feedback loops.The laser signal is converted into an electronic signal by a photodetector and amplified before being added to the injection current of the laser.\\
~\\
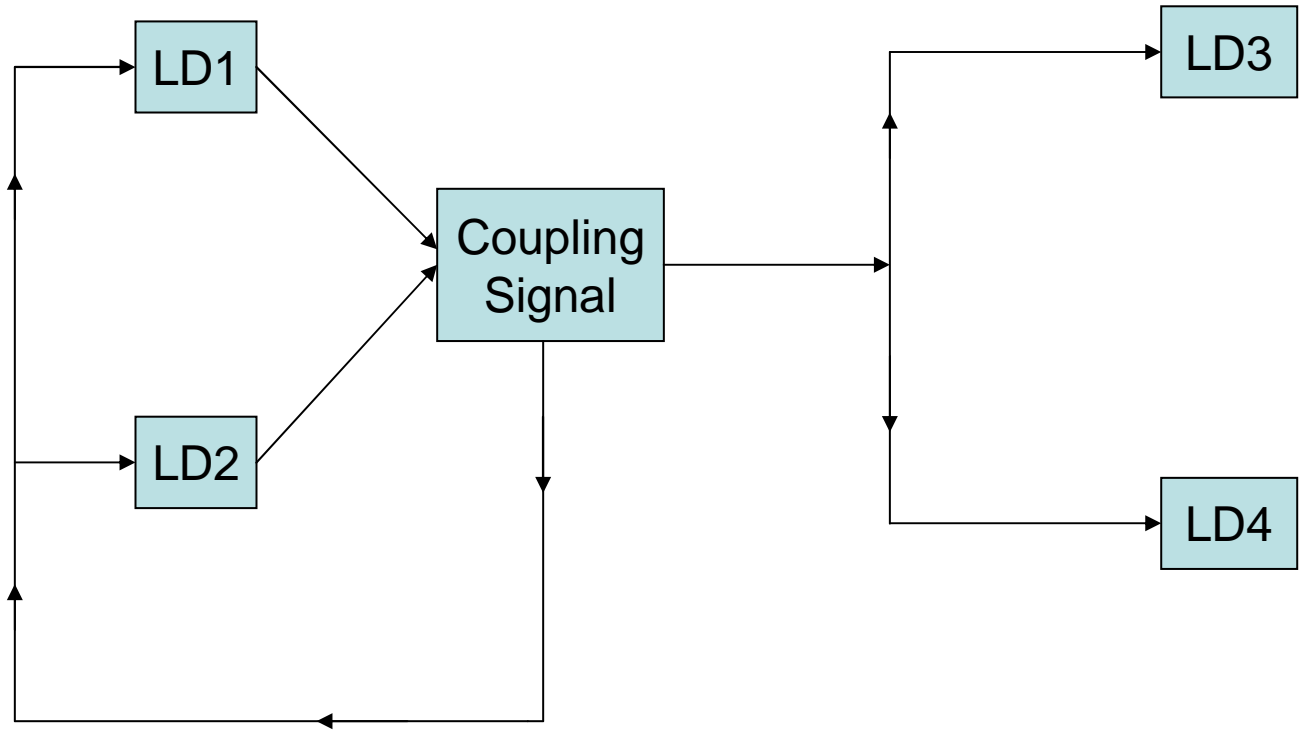
FIG.2.Globally coupled multiplex the transmitter and the receiver systems.LD1 and LD2 are the subsystem laser diodes for the transmitter;LD3 and LD4 are the subsystem laser diodes for the receiver; The coupling signal is fed back to both the transmitter subsystem lasers and added to both the receiver subsystem lasers.\\
~\\
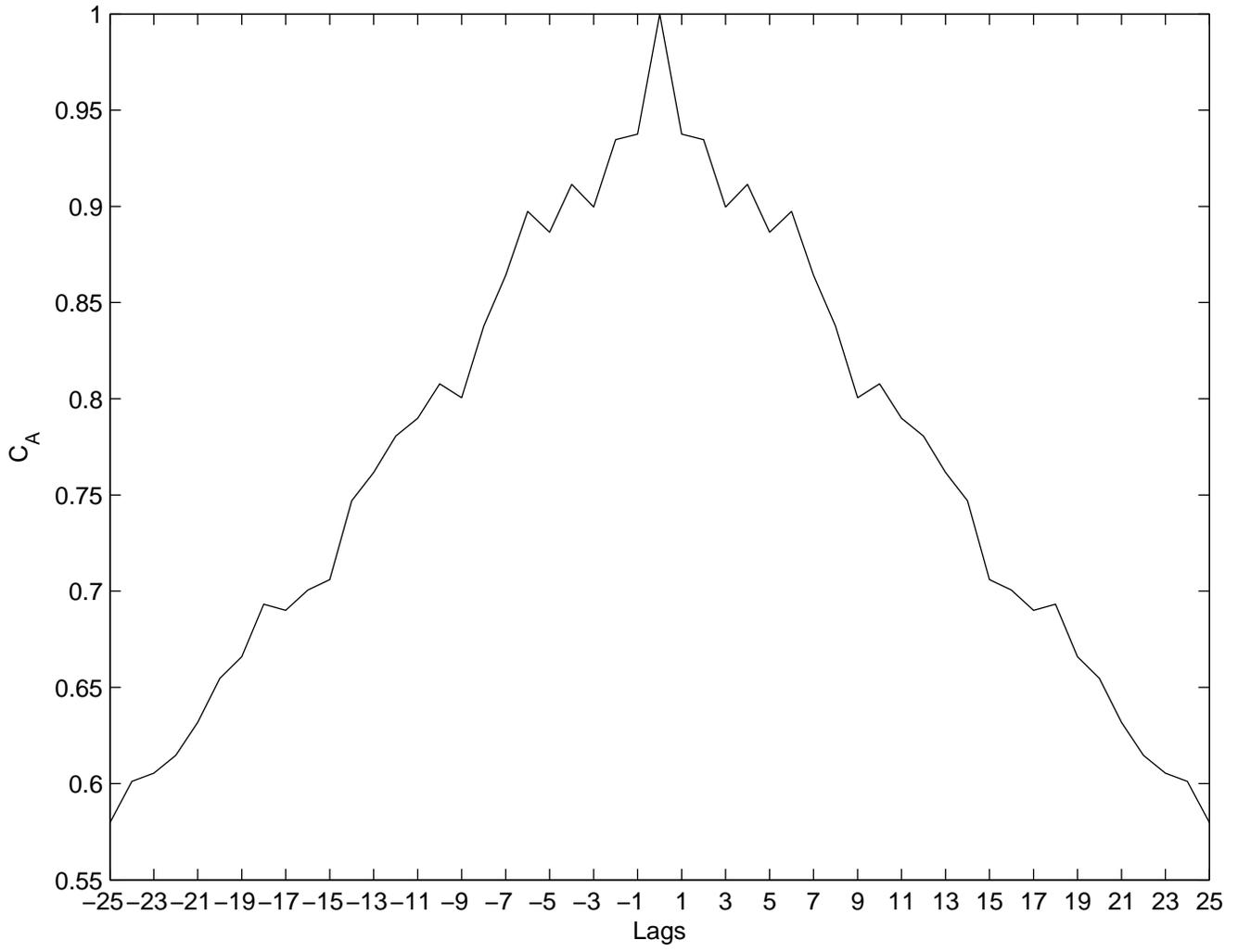
FIG.3. The autocorrelation coefficient $C_{A}$ of the laser output for constant time delays, Eqs.(1-2) for 
$\tau_{01}=4ns, \tau_{02}=7ns,  k_{1}=0.80, k_{2}=0.70.$ Lags are in ns.\\
~\\
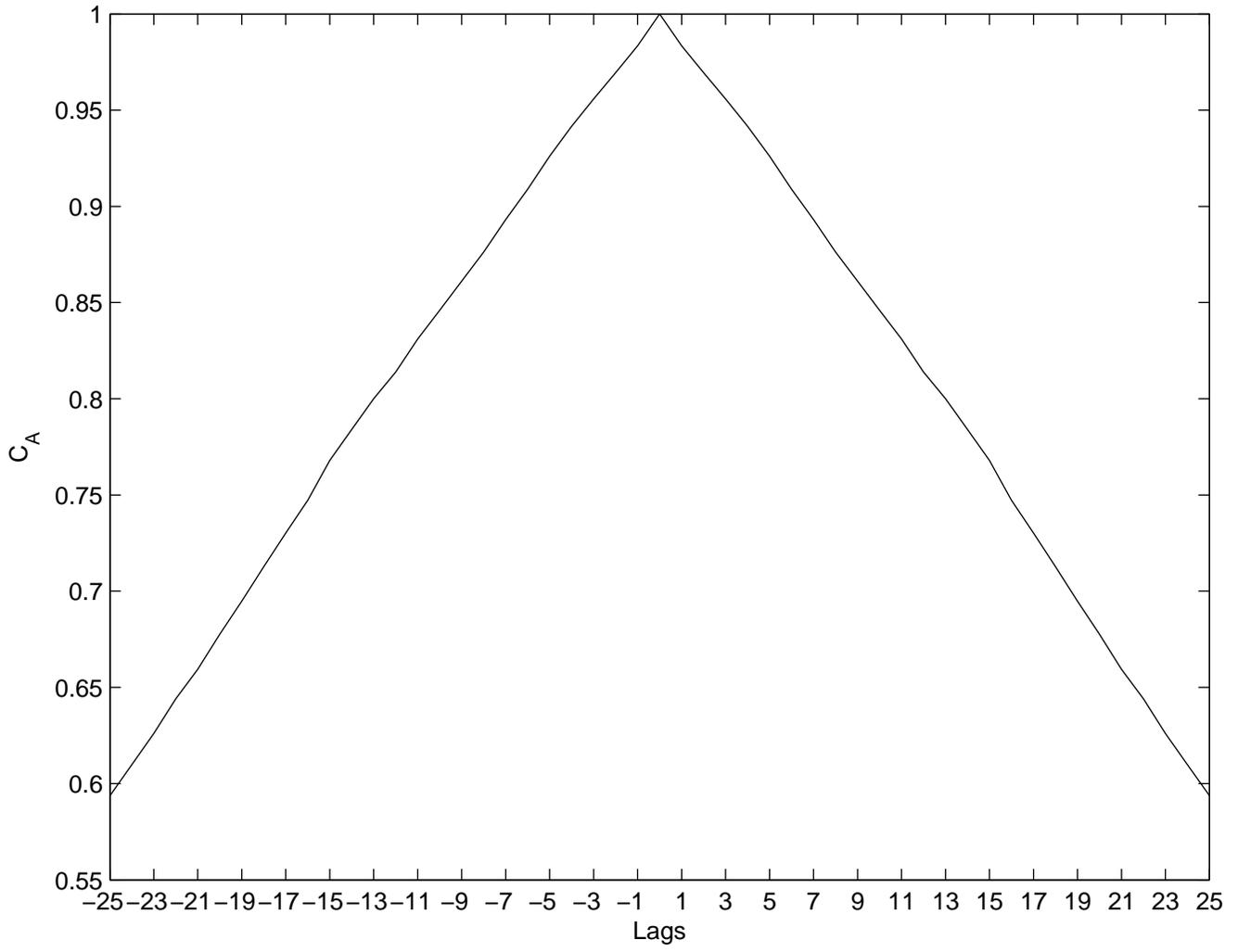
FIG.4. The autocorrelation coefficient $C_{A}$ of the laser output for the product of the sinusoidal and chaotic modulations of time delays, Eqs.(1-2)($K_{1}=0$) for $\tau_{1}(t)= (4X10^{-9}  +  4X10^{-17}S^{f}_{1}(t)\sin(2X10^{6}t))$s and $\tau_{2}(t)= (6X10^{-9}  +  4X10^{-17}S^{f}_{1}(t)\sin(2X10^{6}t))s.$ The other parameters are as in figure 3. Lags are in ns.\\
~\\
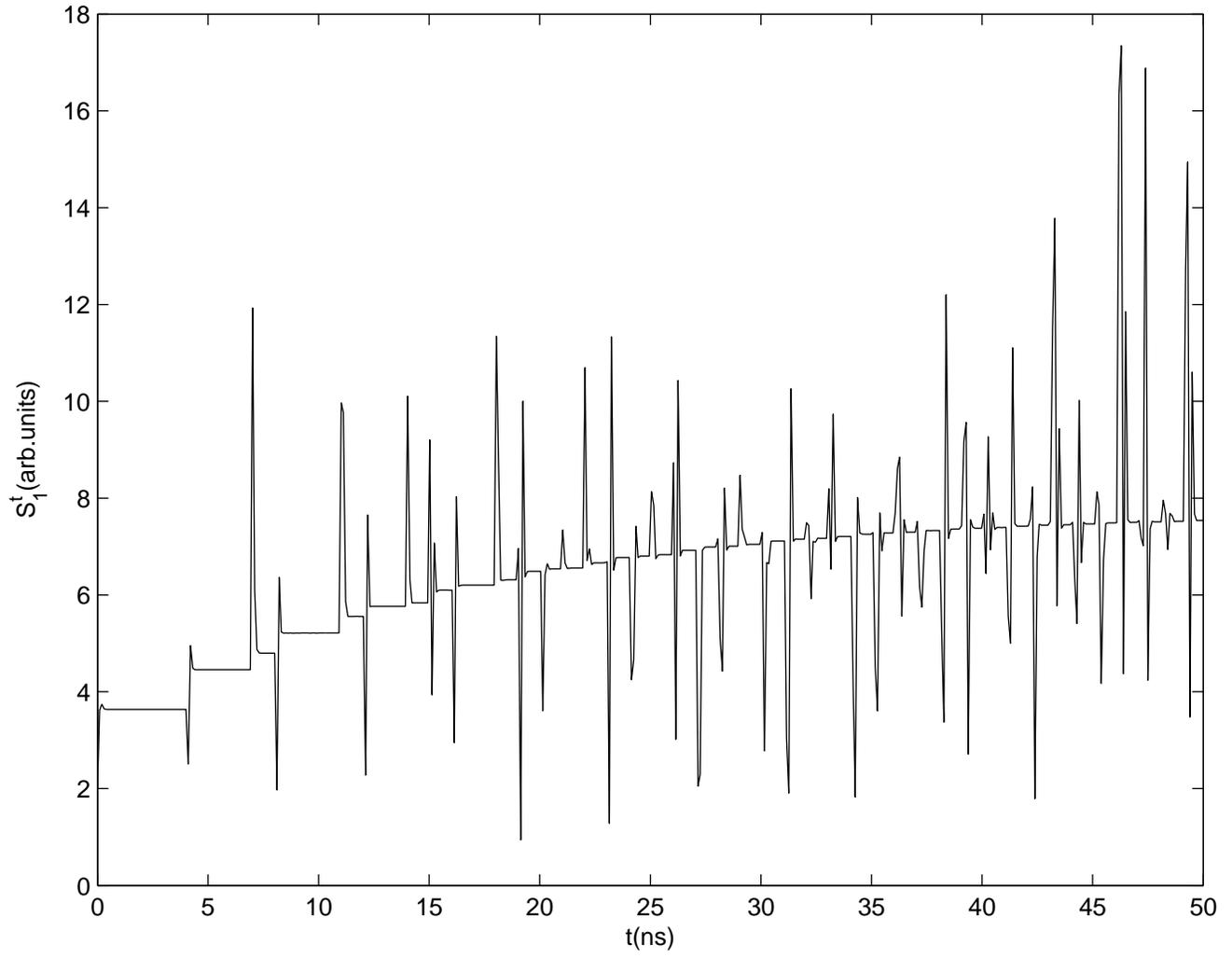
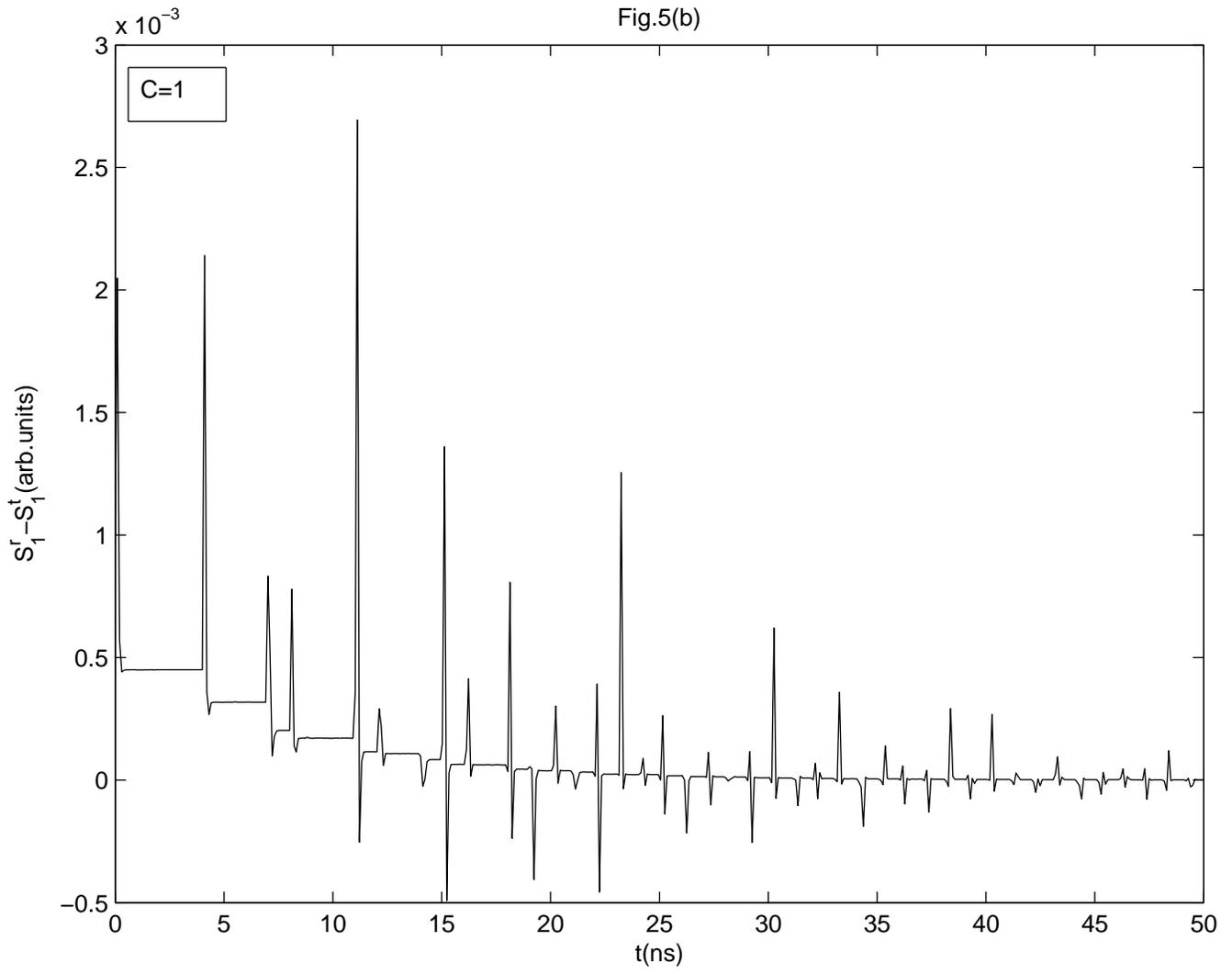
FIG.5. Numerical simulation of globally coupled multiple variable time delay lasers with optoelectronic feedbacks, Eqs.(1-4) for 
$\tau_{1}(t)= (4X10^{-9}  +  4X10^{-17}S^{f}_{1}(t)\sin(2X10^{6}t))$s and \\
$\tau_{2}(t)= (6X10^{-9}  +  4X10^{-17}S^{f}_{1}(t)\sin(2X10^{6}t))s,$
$\tau_{3}(t)= (4X10^{-9}  +  4X10^{-17}S^{f}_{1}(t)\sin(2X10^{6}t))s, k_{1}=0.80, k_{2}= 0.70,k_{3}=0.85, k_{4}=0.75.$
The transmitter subsystems are synchronized to the corresponding receiver subsystems with identical parameters.The coupling signal $S_{1}(t-\tau_{3})$ is added to both the transmitter and reciever subsystems with the coupling rate $1.2Xk_{1}$:
(a) time series of the transmitter subsystem laser intensity $S^{t}_{1}$;(b)Synchronization error dynamics between the receiver subsystem 
$S^{r}_{1}$ and the transmitter subsystem $S^{t}_{1}$ intensities 
$S^{r}_{1}-S^{t}_{1}$. C is the cross-correlation coefficient between the intensities of the transmitter and receiver subsystem lasers.\\
~\\
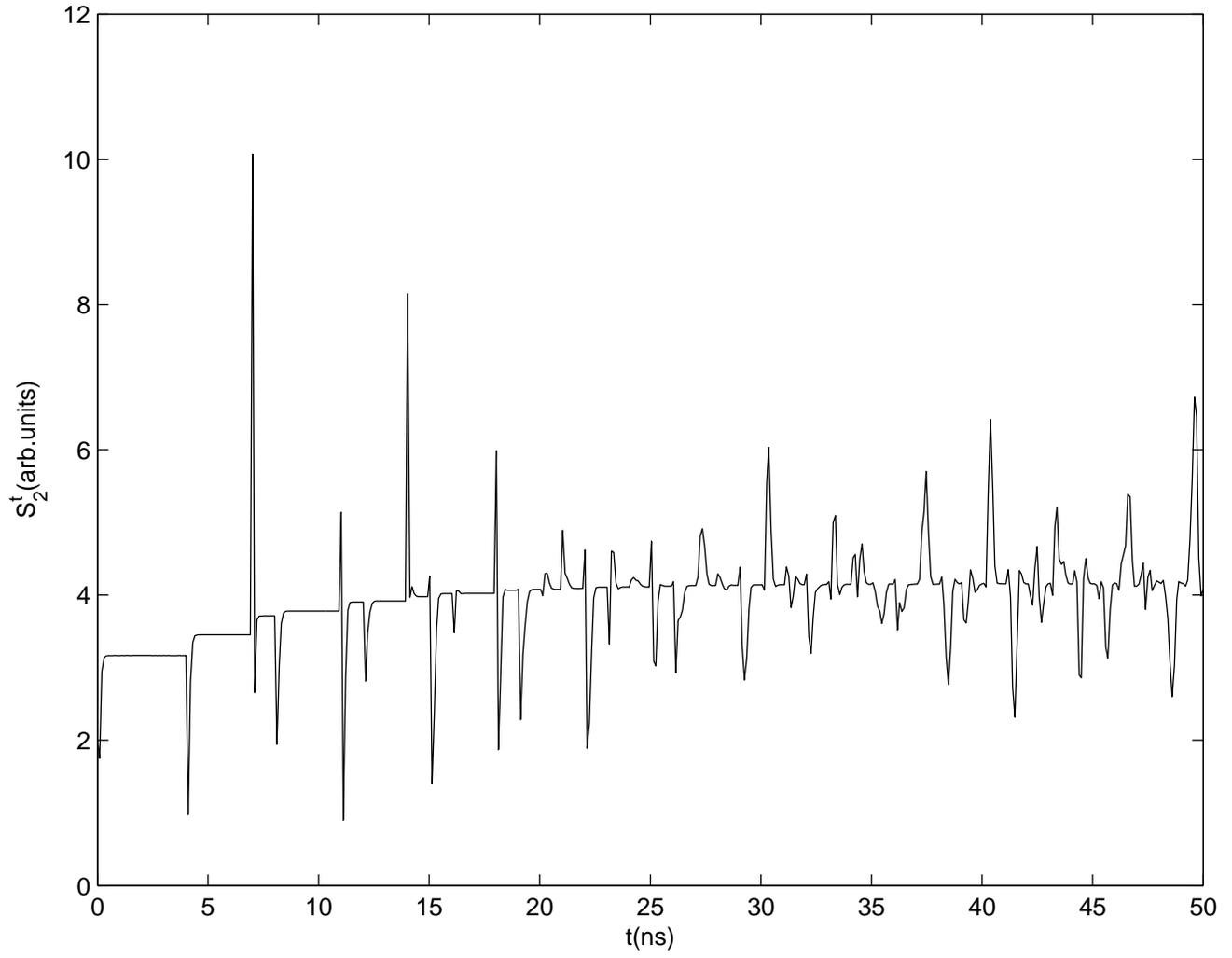
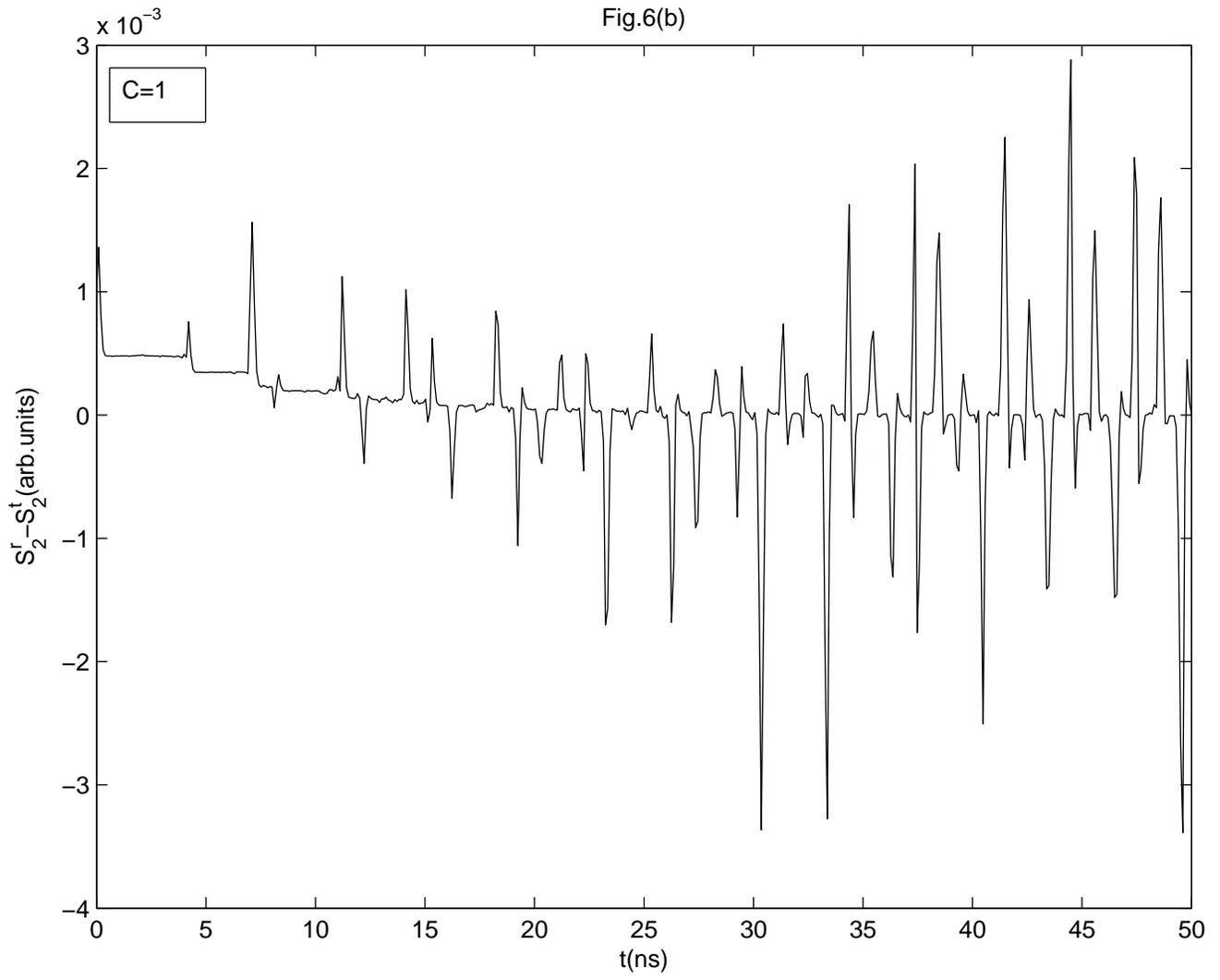
FIG.6.Numerical simulation og globally coupled multiple variable time delay lasers with optoelectronic feedbacks, Eqs.(1-4) for 
$\tau_{1}(t)= (4X10^{-9}  +  4X10^{-17}S^{f}_{1}(t)\sin(2X10^{6}t))$s and \\
$\tau_{2}(t)= (6X10^{-9}  +  4X10^{-17}S^{f}_{1}(t)\sin(2X10^{6}t))s,$
$\tau_{3}(t)= (4X10^{-9}  +  4X10^{-17}S^{f}_{1}(t)\sin(2X10^{6}t))s, k_{1}=0.80, k_{2}= 0.70,k_{3}=0.85, k_{4}=0.75.$
The transmitter subsystems are synchronized to the corresponding receiver subsystems with identical parameters. The coupling signal 
$S_{1}(t-\tau_{3})- S_{2}(t-\tau_{3})$ is added to both the transmitter and reciever subsystems with the coupling rate $0.8Xk_{1}$:(a) time series of the transmitter subsystem laser intensity $S^{t}_{2}$; (b) Error $S^{r}_{2}-S^{t}_{2}$ dynamics. C is the cross-correlation coefficient between the intensities of the transmitter and receiver subsystem lasers.\\

\end{document}